# DEPENDENCE OF SOLAR PLASMA FLOWS ON MAGNETIC FIELD OBLIQUITY

E.J. Zita and C. Smith (The Evergreen State College, Olympia WA, 98505)
N.E. Hurlburt (Lockheed Martin ATC, Palo Alto CA, 94304)

## ABSTRACT

Interactions between flows and magnetic fields in the Sun's plasma (1) can change surface waves and flows near active regions, (2) are evident in cyclic changes of large-scale phenomena such as the meridional circulation, and (3) contribute to dynamics in the long-term solar magnetic cycle, e.g. during the recent prolonged solar minimum 23/24. We investigate possible relationships between these phenomena.

We have observed changes in solar surface flow patterns in active regions, dependent on magnetic field strength and orientation, consistent with the theoretically predicted "Proctor Effect." Other researchers have observed relationships between changes in solar magnetic fields and meridional circulation flows. We explore similarities between the Proctor Effect and the observed interdependence of larger-scale magnetic fields and flows. This may contribute to understanding of fundamental solar convection and dynamo processes, e.g. the prolonged magnetic minimum of the most recent cycle.

## INTRODUCTION

Magnetohydrodynamic (MHD) theory describes a rich variety of behaviors, speeds, orientations, and natures of waves and oscillations in magnetic fields, in plasmas such as those in the Sun. Distinct waves and oscillations can be predicted in regions with specified $\beta$ (gas pressure /magnetic field pressure), propagation vector $\mathbf{k}$ of waves with respect to the magnetic field, inclination angle $\phi$ of the magnetic field with respect to gravity (which is distinct from $\mathbf{k}$), etc. (Thomas 1983, Bogdan et al. 2002, Boyd & Sanderson 2003). While MHD theory often has well-defined wave solutions within given parameter ranges, real plasmas tend to have parameters which vary in space and time, often over different scales in different regions, and waves and flows may transform their characteristics in space and time. For example, the inclination angle $\phi$ of the magnetic field with respect to the vertical varies from 0 (or $\pi$) in sunspot umbrae to $\pi/2$ (horizontal) in quiet regions. Field inclinations can vary significantly throughout active regions, strongly impacting theoretical solutions and actual transformations in waves and flows.

MHD modeling is useful for exploring numerical experiments with more complex situations than can be solved analytically. Detailed measurements from solar observatories permit testing of numerical predictions. Careful analysis of simultaneous datasets on velocities and magnetic fields can reveal interrelations between magnetic fields and velocity fields, as parameters such as $\phi$ and field strength change in a given region. The "Proctor Effect" predicts specific changes in the nature and speeds of flows and waves in magnetoconvection, in response to variations in magnetic field orientation and strength. We investigate connections between predicted and observed photospheric velocity fields, and observed changes in solar velocity patterns during the changing solar magnetic cycle. Deeper understanding of (magneto)convection in regions of changing magnetic fields may illuminate solar magnetohydrodynamics on larger scales.

## NUMERICAL SIMULATIONS AND PREDICTIONS

The Proctor Effect was predicted in MHD models of magnetoconvection in stratified atmospheres with varying inclination angles $\phi$ of the magnetic field with respect to the vertical. Vertical magnetic fields have both steady and traveling solutions, but inclined fields break horizontal symmetry, theoretically permitting only travelling waves (Proctor 1992, Mathews 1992, Thompson 2005). 2D simulations of nonlinear magnetoconvection showed traveling roll solutions in the presence of inclined magnetic fields (Hurlburt, Mathews & Proctor 1996, hereafter HMP). 3D numerical solutions showed convection with a travelling cellular pattern for more vertical fields, and roll-like solutions for more horizontal fields (Hurlburt, Mathews & Rucklidge 2000, hereafter HMR). Recent modeling has confirmed and extended these results under realistic solar conditions (Kitiashvili et al. 2009).

As the mean inclination of a magnetic field through an incompressible fluid layer increases, e.g., as the magnetic field becomes more nearly parallel to the photosphere, 1) the pattern of magnetoconvection cells travels at an increasing horizontal wave velocity, $v_p$, and 2) increasing shear flows produce mean surface flows $\bar{u}$ that can be comparable to the convective flow velocities, $u$ (Fig.1, from HMP). In regions with magnetic fields at steeper inclination to surface flows ($\phi \ll \pi/2$) however, both flows and waves are slower. Where the magnetic field dominates over convection, these waves may travel in the opposite direction to local flows.

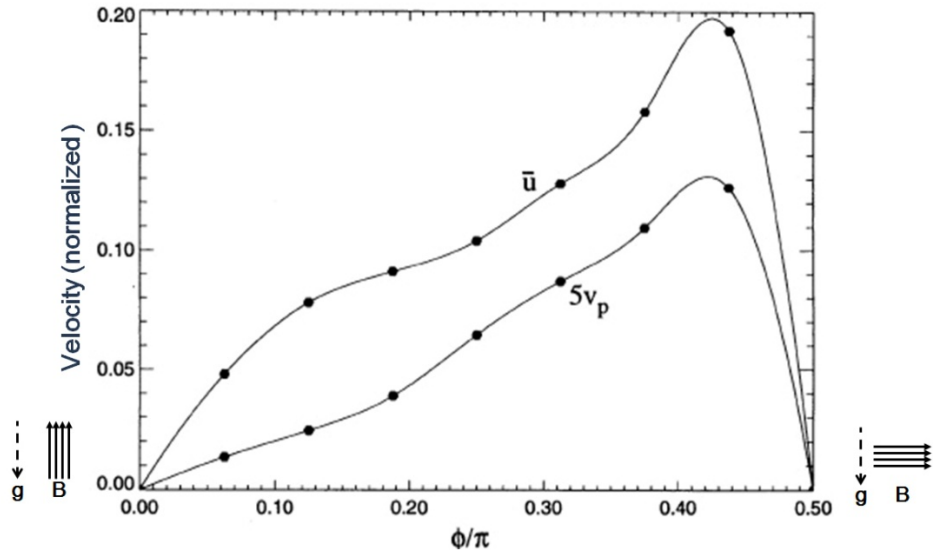

**Fig. 1**. Theoretical wave velocity $v_p$ and mean flow velocity $\bar{u}$ vs. magnetic field inclination angle $\phi$, in a magnetoconvection region with $\beta = 1655$. Mean flows increase as fields become more parallel, at a rate of about 2 m/s per degree of magnetic field inclination (HSZ 2011). Wave speeds are an order of magnitude lower than mean flows (normalized to sound speed; from Fig.5 of HMP).

An important result of HMP's study was that inclined magnetic fields generate horizontal velocities, in addition to the wave velocities already explained by the original magneto-convection. The additional horizontal velocities could not be generated in the simulation without oblique magnetic fields. The velocity vs. magnetic field obliquity relation evident in Fig.1 can be quantified using the simulation's normalization to the photospheric sound speed, $c = 7$ km/s.

This yields a decrease in speeds of about 2 m/s per degree of obliquity in the magnetic field. HMP 1996 concluded with a call for investigation of such behavior in real solar data.

## METHODOLOGY

We tested theoretical Proctor Effect predictions by analyzing high resolution velocity and magnetic field data in solar active regions (Smith, Zita & Hurlburt 2010, hereafter SZH), and comparing observations to HMP's numerical simulations of nonlinear, compressible convection in oblique magnetic fields.

We analyzed data from the Solar Optical Telescope (SOT) on the Hinode[1] Spacecraft, recovering plasma velocities from filtergrams, and magnetic field inclinations and strengths from spectro-polarimeter data. Although the field of view is much smaller than the Solar Dynamics Observatory[2] (SDO), which looks at the entire solar disc, Hinode remains the highest resolution instrument observing the Sun.

We obtained and aligned velocity fields ($v$) and magnetic field vectors ($B$) from Hinode data near active regions. Filtergram (FG) images in the G-Band, also from Hinode's SOT, were used to calculate horizontal velocities of the plasma in the photosphere. First we calibrated the images with the SolarSoftWare (SSW) package for processing with Interactive Data Language (IDL), to remove artifacts and correct for Hinode instrumental idiosyncrasies. To obtain velocity fields from FG data, we then applied the LMSAL local correlation tracking tool (LCT) written for IDL (Hurlburt, Shine & Simon 1995) to G-band datasets using a 32x32 grid on the 1024x1024 images. The calculated flow vectors were then averaged over the time interval in which both FG and SP data were recorded (one hour). The resultant 2D vector fields are shown in Fig.2a (for an active region) and Fig.3a (for a quiet region).

The corresponding Hinode vector magnetograms were aligned and binned to a comparable resolution. We used data from SOT's spectropolarimeter (SP) for magnetic field inclination, strength, and azimuth. Magnetic field strength is highest in the central, vertical field of an active region (Fig.2b), and while inclination angle is not disambiguated, antiparallel field orientations can be clearly distinguished in Fig.2c. In quiet regions (Fig.3) the magnetic field strength is weaker and more uniform, and inclination angles are more horizontal in quiet regions.

Finally, we used IDL to plot velocities calculated with our LCT against magnetic field strength and inclination $\phi$. In order to compare SP and FG data to each other, they were cropped, smoothed, and aligned so that they had the same field of view, size, and scale; and we corrected for drift.

---

[1] Hinode is a Japanese mission developed and launched by ISAS/JAXA (Sept. 2006), collaborating with NAOJ, NASA and STFC (UK), ESA, and NSC (Norway).
[2] SDO is a NASA mission developed collaboratively by Lockheed Martin Solar Astrophysics Laboratory (AIA), Stanford University (HMI), and University of Colorado (EVE), launched in Feb. 2010.

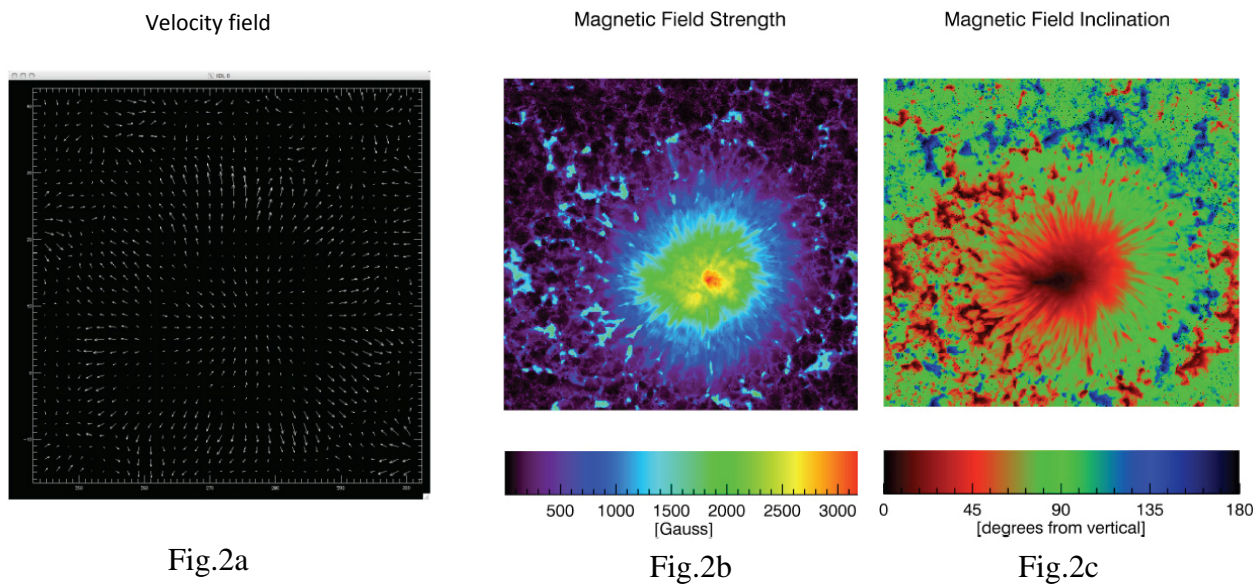

Fig.2a     Fig.2b     Fig.2c

**Fig.2**: Velocity field (a), magnetic field strength (b), and magnetic field inclination (c), in a magnetic active region, AR 10944 (1 Mar.2007) (from SZH 2010).

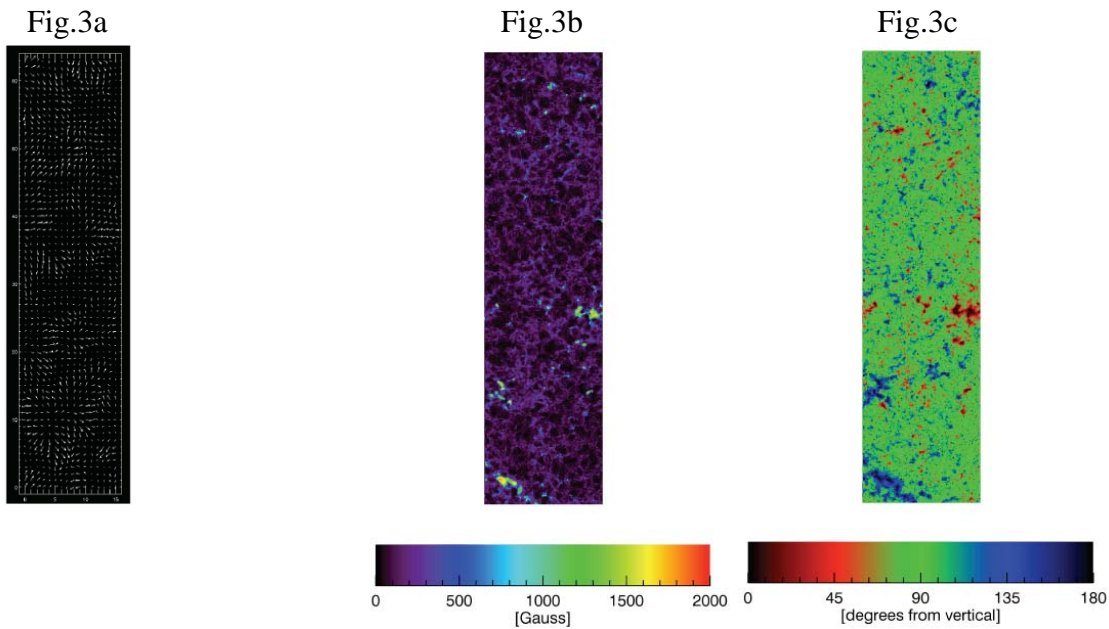

Fig.3a     Fig.3b     Fig.3c

**Fig.3**: Same as Fig.2, except in a region of "quiet Sun" with more uniform, weaker magnetic field.

# RESULTS

**Results in Active Regions**: We show observed velocities *v* versus magnetic field inclination angles $\phi$ near active regions in Fig.4a, and overlay our data with theoretical $v(\phi)$ relations. We plot $v(B)$ in Fig.4b in order to understand scatter in the data. Our analysis does not distinguish surface waves from phase velocities, as computations of HMP do. This information could be recovered in future work. Error bars are omitted from data points, for clarity. Simulations are from a slightly deeper layer in the convection zone than observations, which are taken at the photosphere.

Our findings (SHZ 2010) are generally consistent with the Proctor Effect predictions (HMR 1996), qualitatively and quantitatively. A representative dataset is illustrated by AR 10944 (1 Mar.2007). In Fig.4a, observed photospheric velocities increase as the magnetic obliquity angle increases, that is, as fields become more parallel to the surface. Upper speed limits in Fig.4a suggest an inclination dependence similar to Proctor Effect predictions for mean flows, $\bar{u}$ (Fig.1), with a velocity/inclination relationship of roughly 2 m/s per degree. Velocities rapidly decrease in the perfectly perpendicular and parallel limits, as in the Proctor Effect. A preponderance of observed velocities up to an order of magnitude lower than mean flows $\bar{u}$, in Fig.4a, is also consistent with predictions for magnetoconvection waves, $v_p$.

**Fig.4a** (left): Velocities vs. inclination angle $\phi$. Solid lines: Proctor Effect predictions from model of HMP 1996 (see Fig.1). Scatter plot: Data from AR 10944 (1 Mar.2007). Observed surface velocities increase roughly linearly with magnetic field obliquity angle, for $\phi < 0.45\,\pi$. (SZH 2010)

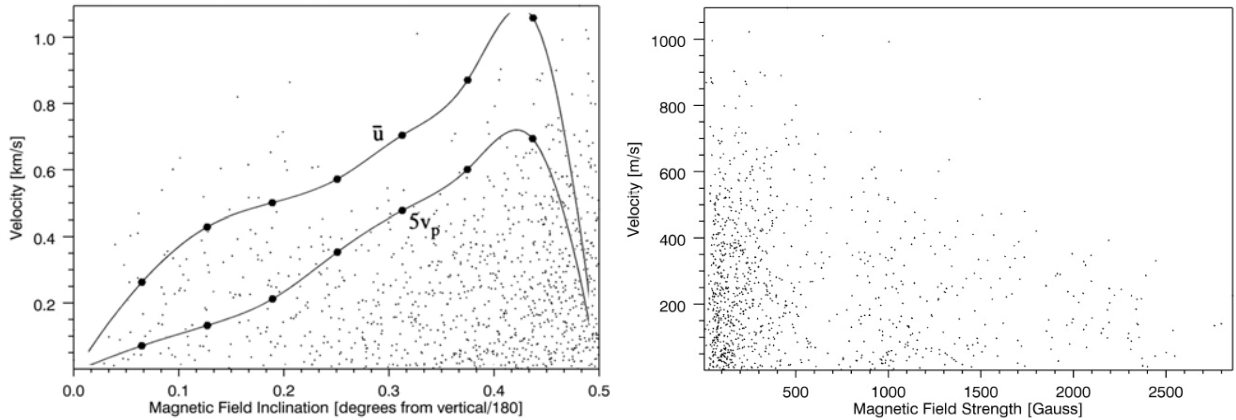

**Fig. 4b** (right): Observed velocities vs. magnetic field strength in active region. Speeds generally decrease in regions with stronger magnetic fields. (SZH 2010)

Observed velocities vs. local field inclinations $v(\phi)$ show significant scatter (Fig.4a). We investigated the source of this scatter by plotting velocity versus magnetic field strength. Fig.4b reveals that observed flows are faster in weak field regions, and that velocities are lower in regions with stronger magnetic fields. This is consistent with magnetic resistance to flows in conducting media. Sorting the points by field strength (not shown) indicates that this inverse relationship between *v* and *B* (Fig.4b) correlates with scatter in the $v(\phi)$ plot (Fig.4a). Similarly, the linear relationship between *v* and $\phi$ (Fig.4a) correlates with scatter in $v(B)$ in Fig.4b.

**Results in quiet region**: Velocities are slightly higher where fields are weaker, more homogeneous, and generally more horizontal.  Most data points are for weak fields here (and none are near poles).  Patterns are similar to those in active regions.  Flows show a strong preference for more parallel field regions, with no clear drop-off at $\phi = 90°$.

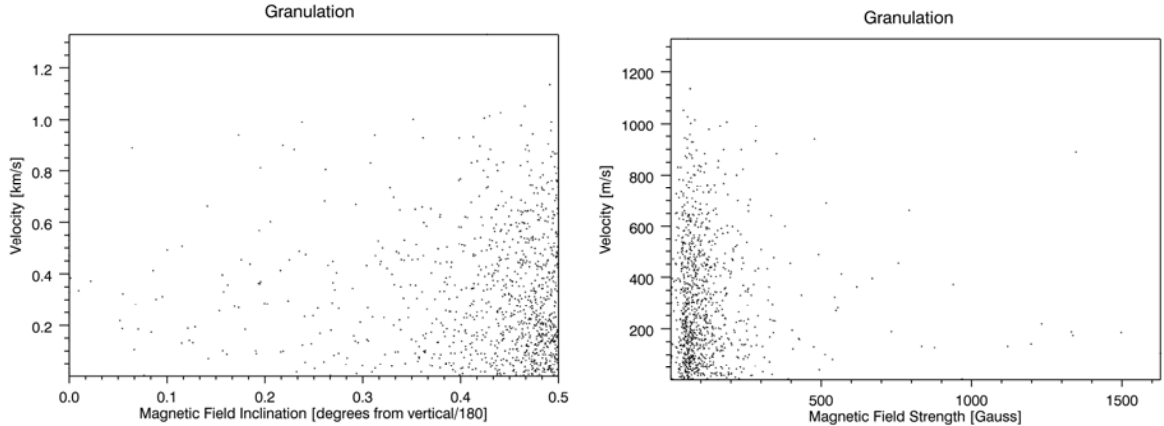

Fig.5a(left) and Fig.5b (right):  $v(\phi)$ and $v(B)$ in quiet region corresponding to Fig.3  (SZH 2010).

Despite significant scatter, some general trends are evident in the data. For a given magnetic field strength, we find that waves and flows are faster in fields which are oriented more nearly horizontally, and slower in magnetic fields tilted more vertically, whether into or out of the surface of the Sun.  These results are consistent with the Proctor Effect, with an inclination relationship of approximately 2 m/s per degree of field tilt.  For a given magnetic field obliquity, velocities are lower in regions of stronger magnetic field, and can be higher in regions with weaker field.  These results are consistent with the fundamental understanding that high field strengths and concentrations inhibit solar convection and thus reduce horizontal surface velocities.

### MERIDIONAL CIRCULATION AND MAGNETOCONVECTION

Magnetoconvection is convection threaded with magnetic flux:  transport of energy by fluid motion, where flows are modified by the magnetic Lorentz force, and flows move fields by the induction equation (Stein 2012).  Mass conservation determines topology, and the scale of magnetoconvection is set by boundary conditions (e.g. the efficiency of radiative cooling).  Magnetoconvection has traditionally been used to describe cells near sunspots, where density stratification is a key factor.  While high $\beta$ weakens the effects of fields on flows deeper in the convection zone, meridional circulation can also fit the general description of magneto-convection.  These two phenomena have different driving mechanisms and operate on different scales.  Temperature gradients are a major driver of small-scale magnetoconvection in solar active regions.   Meridional circulation in the solar convection zone is driven by angular momentum transport and the Coriolis force (Miesch & Hindman 2011).

We investigate observed changes in fields and flows in meridional circulation across solar cycles, and compare these to the Proctor Effect. Helioseismologic observations reveal poleward flows at maximum speeds of about 20 m/s, with surface flows taking about 11 years from equator to poles; by mass conservation, return flows from poles to equator are inferred along the tachocline, taking about 22 years (Basu & Antia 1997, Komm et al. 2012). While the direction of this meridional circulation $u_{MC}$ is generally consistent, its relative orientation with respect to solar $B$ varies as the magnetic field changes during the solar cycle. The meridional circulation tends to be generally parallel (or antiparallel) to the local dipolar field of the quiet Sun at midlatitudes and at poles, except that fields and flows tend to be mutually perpendicular in the equatorial return region, and oblique at high latitudes. The active Sun, on the other hand, can have many multipolar magnetic field regions, with components oblique or perpendicular to the meridional circulation at midlatitudes as well. Magnetic field strengths vary considerably during the solar cycle, from a smooth, strong mean amplitude at solar minimum to blotchy strong magnetic regions during peak activity, when the mean field is relatively weak.

The speed of the meridional circulation varies with the phase of the solar cycle. Hathaway & Rightmire (2010, hereafter H&R) and Komm et al. (2012) found that the meridional circulation is faster at solar minimum, when velocities are more parallel to magnetic fields, and slower at solar maximum, when much more magnetic flux is oblique and perpendicular to surface flows.

Qualitatively, these observations are consistent with the Proctor Effect: magnetoconvective flows and waves can be faster where they are more nearly parallel to magnetic fields, and slower in stronger and more perpendicular field regions. (The effect vanishes where magnetic fields are perfectly parallel to flows, e.g. at the poles, and where they are perfectly perpendicular, e.g. at equatorial return regions.)

**Quantitative effects of fields on flows**: We next review rates of changes of flows of meridional circulation observed during the solar cycle. By tracking magnetic elements from MDI on SOHO[3], H&R (2010) found that the dominant, dipolar component (the $l = 2$, $m = 1$ Legendre polynomial) of the meridional circulation's flow amplitude nearly doubled after cycle 23, from about 8 m/s at solar max to about 14 m/s at solar min (Fig.4 H&R 2010). Using ring diagram analysis from HMI on SDO[4] (and literature review), Komm et al. (2012) found that the maximum meridional flow is about 20 m/s, the flow amplitude increases about 5 m/s from solar maximum to solar minimum, and the magnetic tracking technique yields speeds about 40% too low. This does not necessarily negate H&R's evidence of an unusual variation in meridional circulation from solar cycle 23 to 24; it would raise their estimates to about 11 m/s and 20 m/s, an increase of 9 m/s in the recent prolonged solar minimum.

Meridional circulation speeds peak at midlatitudes (Fig.3 H&R 2010, Fig.7 Komm et al. 2012), where flows are parallel to the solar surface. At solar minimum, these flows are nearly parallel to mean dipolar magnetic fields. At solar maximum, multipole moments increase the proportion of magnetic fields with oblique components to the flow. The angles of obliquity vary continually in time and space, and the density of oblique fields varies with the filling factors of multipoles, as active regions come and go, and reversed flux rises and reconnects.

---

[3] Michelson Doppler Imager (MDI) on the European Space Agency (ESA)/National Aeronautics and Space Administration (NASA) Solar and Heliospheric Observatory (SOHO)

[4] Helioseismic and Magnetic Imager (HMI) instrument on the Solar Dynamics Observatory (SDO)

We have seen that speeds of mean flows in magnetoconvection simulations increase at a rate of 2 m/s per degree (Fig.1, Fig.4a) as magnetic field orientation becomes less oblique, by the Proctor Effect. If we estimate the averaged photospheric magnetic field obliquity at solar max as $\phi = 5°$ greater than that at solar minimum, then this effect yields changes in surface flow speeds of 10 m/s, the same order of magnitude as the changes in meridional flow speeds observed by both H&R 2010 and Komm et al. 2012.

We have shown that changes in meridional circulation flow speeds with respect to mean magnetic fields during the solar cycle are consistent, qualitatively and quantitatively, with the Proctor Effect, we have argued that this effect could plausibly account for observed changes in meridional circulation speeds through the solar cycle.

### SUMMARY AND FUTURE WORK:

We studied Hinode data near active regions and found that observed surface velocities (SZH 2010) behave with respect to magnetic field inclinations consistently with theoretically predicted Proctor effects in magnetoconvection simulations (HMP 1996). This effect is evident in active regions: magnetoconvection flows and waves are observed to propagate more slowly in regions with oblique magnetic fields, and faster in regions with more nearly parallel magnetic fields (with velocities dropping off in the parallel and perpendicular field limits). Observed velocities increase by about 2 m/s per degree of magnetic field obliquity decrease (Fig.4a). Quiet regions show similar trends, with slightly higher speeds (and no clear dropoff in the parallel field limit).

This effect might help explain the observations of increased meridional circulation speeds during extreme magnetic minima, when surface magnetic fields are far less oblique (H&R 2010). Observations of changes in flow speeds of meridional circulation during the solar cycle (H&R 2010, Komm et al. 2012) are consistent with the Proctor Effect, both qualitatively and quantitatively. Both of these phenomena can occur in relatively high-$\beta$ convection zone plasma, as perturbations on primary drivers (e.g. temperature gradients or angular momentum transport). Mean flows can be decelerated by oblique surface magnetic fields, or sped by more parallel fields.

We summarize how the Proctor Effect might affect the meridional circulation's flow speed. Fundamentally, the meridional circulation is driven by angular momentum transport and the Coriolis force; factors such as thermal gradients and meridional momentum transport may also play a role (Miesch & Hindman 2011). Then, its speed may be modified throughout the solar cycle by factors including changing orientations of magnetic fields. Approaching solar maximum, more active regions and magnetic dipoles rise to the solar surface, due to solar dynamo processes. The resultant oblique magnetic fields can impede the flow of the meridional circulation, by the Proctor Effect (and perhaps by other mechanisms as well). Relaxing toward solar minimum, near-surface fields become more parallel, and flows regain speed. The dominant component of meridional circulation may increase from speeds of about 11 m/s at solar max to 20 m/s at solar min (H&R 2010, Komm et al. 2012). This would correspond to an increase in average magnetic field obliquity of $\phi < 5°$ (from HMP 1996 and SZH 2010). Behavior of flows near the tachocline is beyond the scope of this paper.

The phenomena discussed above are only a subset of the rich variety of interactions between flows and magnetic fields in the Sun. As plasma $\beta$ increases with depth beneath the photosphere (Weiss et al. 1990), nonlinear effects must be considered. Not only do fields modify flows, but flows also modify fields. In particular, nonlinear interactions between plasma flows and magnetic fields drive the solar dynamo in the convection zone (Weiss & Thompson 2009, Charbonneau 2010, Miesch 2012). While the Proctor Effect may make a significant contribution, it can be only one of many aspects of solar convection zone dynamics.

The next step in testing our idea would be to directly investigate the average inclination of surface magnetic fields leading up to the recent solar cycle. Did multipoles relax unusually quickly after solar maximum at the end of cycle 23, leaving a smooth magnetic dipole in place unusually early? If so, did this facilitate faster meridional circulation during the solar minimum before cycle 24, interfering with the usual tachocline storage and dynamo amplification of magnetic flux? Could this be an example of Proctor Effects contributing to the modification of meridional circulation and the solar dynamo? We propose such a test for future investigation.

This work was supported by NSF grant 0807651, NASA grants NAS5-38099, NNM07AA01C, NNG04EA00C, and Lockheed Martin Internal Research Funds. We thank colleagues R. Shine, M. Cheung, T. Berger, M. DeRosa, A. Sainz-Dalda, G. Slater, Z. Frank and R. Seguin at Lockheed Martin Solar and Astrophysics Laboratory (LMSAL) for generously providing guidance and facilities for this research.

### REFERENCES:

Basu, S., Antia, H.M. 1997 Seismic measurement of the depth of the solar convection zone, *MNRAS* **287**, 189

Berger, T. et al. 2009 March 09, Hinode Solar Optical Telescope Data Analysis Guide, Ver.3.3 LMSAL internal publication

Bogdan, T. J., Rosenthal, C. S., Carlsson, M.; Hansteen, V., McMurry, A., Zita, E. J., Johnson, M., Petty-Powell, S., McIntosh, S. W., Nordlund, Å., Stein, R. F., Dorch, S. B. F. 2002 Waves in magnetic flux concentrations: The critical role of mode mixing and interference, *Astronomische Nachrichten*, **323**, 196 (doi: 10.1002/1521-3994(200208)323:3/4<196::AID-ASNA196>3.0.CO;2-E)

Boyd, T.J.M., and Sanderson, J.J. 2003 The Physics of Plasmas, Cambridge University Press

Charbonneau, P. 2010 Dynamo Models of the Solar Cycle. *Living Reviews in Solar Physics*, **7,** 3 (doi:10.12942/lrsp-2010-3)

Dikpati, M. & Charbonneau, P. 1999, A Babcock-Leighton Flux Transport Dynamo with Solar-like Differential Rotation. *Astrophys. J.* **518**, 508 (doi: 10.1086/307269)

Hathaway, D.H. & Rightmire, L. 2010 Variations in the Sun's Meridional Flow over a Solar Cycle. *Science* **327**, 1350-1352 (doi: 10.1126/science.1181990)

Hurlburt, N.E., Matthews, P.C., Proctor, M.R.E. 1996 Nonlinear Compressible Convection in Oblique Magnetic Fields. *Astrophys. J.* **457**, 933-938 (doi: 10.1086/176786)

Hurlburt, N. E., Matthews, P. C., Rucklidge, A. M. 2000 Solar Magnetoconvection (Invited Review) *Solar Phys.* **192**, 109 (doi: 10.1023/A:1005239617458)

Kitiashvili, N., Kosovichev, A. G., Wray, A. A., and Mansour, N. N. 2009 Traveling Waves of Magnetoconvection and the Origin of the Evershed Effect in Sunspots. *Astrophys. J.* **700**, L178 (iopscience.iop.org/1538-4357/700/2/L178/)

Komm, R., González Hernández, I. Hill F., Bogart, R., Rabello-Soares, C., Haber D. 2012 Subsurface Meridional Flow from HMI Using the Ring-Diagram Pipeline. *Solar Phys*ics, *Online First* (doi:10.1007/s11207-012-0073-y)

Mackay, D.H. 2012 The Sun's Global Magnetic Field. *Phil. Trans. R. Soc. A,* 370, 536 (doi: 10.1098/rsta.2011.0536)

Miesch, M.S. 2012 The Solar Dynamo. *Phil. Trans. R. Soc. A,* 370, 507 (doi: 10.1098/rsta.2011.0507)

Miesch, M.S. & Hindman, B.W. 2011 Gyroscopic Pumping in the Solar Near-surface Shear Layer. *Astrophys. J.* **743**, 79 (doi: 10.1088/0004-637X/743/1/79)

Proctor, M.R.E. 1992 Magnetoconvection. In J. H. Thomas and N. O. Weiss, eds, *Sunspots: Theory and Observations*, Kluwer, Dordrecht, p.221

Smith, C., Zita, E.J., Hurlburt, N.E. 2010 Solar Plasma Flows and Convection in Oblique Magnetic Fields. *APS-NW* #Ds1.005 (http://labs.adsabs.harvard.edu/ui/abs/2010APS..NWS.D1005S)

Stein, R.F. 2012 Magneto-convection. *Phil. Trans. R. Soc. A,* 370, 533 (doi: 10.1098/rsta.2011.0533)

Thomas, J.H. 1983 Magneto-atmospheric waves. *Ann.Rev.Fluid Mech.* **15**, 321 (doi: 10.1146/annurev.fl.15.010183.001541)

Thompson, S.D. 2005 Magnetoconvection in an inclined magnetic field: Linear and weakly non-linear models. *MNRAS* **360**,1290 (doi: 10.1111/j.1365-2966.2005.09127.x)

Weiss, N. O., Brownjohn, D. P., Hurlburt, N. E., Proctor, M. R. E. 1990 Oscillatory convection in sunspot umbrae. *MNRAS* **245**, 434

Weiss, N. O. & Thompson, M. J. 2009 The Solar Dynamo. *Space Science Reviews*, 144, 53-66 (doi: 10.1007/s11214-008-9435-z)

Zita, E.J., Smith, C., Hurlburt, N.E. 2012 Interdependence of solar plasma flows and magnetic fields. AAS #202.09 (http://labs.adsabs.harvard.edu/ui/abs/2012AAS...22020209Z)

Zita, E.J. 2010 Sensitivity of a Babcock-Leighton Flux-Transport Dynamo to Magnetic Diffusivity Profiles. astroph (arXiv:1009.5965)